\documentclass[pdflatex,sn-mathphys-num]{sn-jnl}
\usepackage{mathrsfs}

\usepackage{graphicx}%
\usepackage{multirow}%
\usepackage{amsmath,amssymb,amsfonts}%
\usepackage{amsthm}%
\usepackage{mathrsfs}%
\usepackage[title]{appendix}%
\usepackage{xcolor}%
\usepackage{textcomp}%
\usepackage{manyfoot}%
\usepackage{booktabs}%
\usepackage{algorithm}%
\usepackage{algorithmicx}%
\usepackage{algpseudocode}%
\usepackage{listings}%
\usepackage{lmodern}%
\usepackage{tabularx}%
\usepackage{multirow}

\theoremstyle{thmstyleone}%

\theoremstyle{thmstyletwo}%

\theoremstyle{thmstylethree}%

\begin{document}

\title[Article Title]{Entropic Approach to Critical Materials Assessment }

\author{\fnm{Alan J.} \sur{Hurd}}\email{ahurd48@msn.com}

\affil{

\orgname{Los Alamos National Laboratory}, \orgaddress{\street{PO Box 1663}, \city{Los Alamos}, \postcode{87545}, \state{NM}, \country{USA}}}

\abstract{
Most methodologies for materials criticality assessment score supply risk and societal importance.  Market-based criteria offer quantitative measures for assessment.  Here we develop a statistical 
approach based on a geologic entropy function in which flexible constraints--such as economic, national security related, or regulatory--can be applied. As an example, the formulation describes the relation between elemental price and crustal abundance for selected elements, both important to supply risk.  The method may be applicable to parameters resulting from collective decisions exhibiting a highly peaked probability distribution.
}

\keywords{critical materials, geologic, supply chain, sustainability, quantum materials, material availability}

\maketitle

\section{Introduction}\label{intro}

As geopolitical events in early 2025 revealed, supply risk for critical materials\footnote{Terms used in this paper for critical entities include, in order of increasing set size, $isotopes \subseteq elements \subseteq minerals \subseteq materials \subseteq commodities$; they need not be stable, non-fuel, crystalline, solid or even a condensed phase. \textit{Critical materials} is used as the generic label.  In this unconventional convention, $isotopes$ include predominantly manmade elements such as gold isotopes (e.g. $^{198}Au), Tc, Pu$, and $Pm$. $Elements$ are single-nucleus chemical elements including their isotopes such as gold ($^{197}Au$), natural $He$ and its isotopes $^{3}He$, and $^{4}He$.  $Minerals$ are naturally occurring elements, compounds, glasses, and mixtures such as gold, mercury, stishovite, and obsidian.  $Materials$ are processed minerals and elements including dental gold alloys, steels, gallium nitride, and heavy water.  $Commodities$ are any element, mineral, or material that is traded in markets such as $^{3}He$, copper, silicon, and gravel.} can surge dramatically.  The tool for assessing supply-chain robustness is criticality assessment (CA) in which both societal importance and supply risk of a chemical element, mineral, or material is evaluated; elements with high importance and risk are the most critical.  Thus, for example, many of the rare earth elements, such as dysprosium, appear in the high-risk, high-importance quadrant owing to their role in energy security for high strength magnets in electric vehicles and generators.  Supply risk is high for $Dy$ and all rare earths due to the near monopoly owned by China in mining and fabricating products.  

The most recent \textit{critical commodities} list developed by the US Geological Survey (USGS) \cite{usgscommodity} in March 2025 has 84 entries about 30 of which are not chemical elements but ores or processed materials.  Among the entries are high-volume materials aluminum, silicon, clays, gravel, and sand; while none of these experience inherent supply risk in the US, they score above the risk threshold owing to their volume and importance.  

The official US \textit{critical minerals} list, separate from the USGS commodities list, is founded on CA methodology \cite{energyact}, \cite{usgsmethodology} and is considered the 'gold-standard' for domestic uses; it was updated by the USGS in 2022 \cite{usgs}. It contains 50 minerals and elements.  Although the US Energy Act of 2020 \cite{energyact} defines "critical" as "essential to the economic or national security of the United States", the scope of the law immediately excludes fuel minerals, such as uranium, thorium, and man-made plutonium. The list also excludes helium and a byproduct of extracting another element unless the byproduct is economically viable.  Not surprisingly isotopes were excluded from the 2022 list.  

\subsection{Entropic Approach to Criticality Assessment}\label{entropy}

A promising way to analyze commodity price with constraints is through a statistical variational method.  In collaboration with R. Perumal Ramasamy of Anna University \cite{perumal}, we developed an expression for (extractive) geologic entropy and used it to explore a family of utility functions that fit extant data well.

\begin{figure}[htbp]
    \centering
    \includegraphics[width=0.7\linewidth]{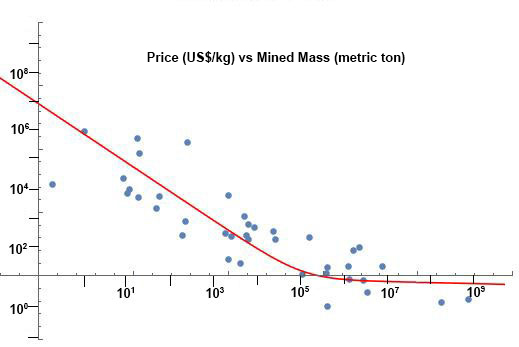}
    \caption{Data of price (US\$/kg) vs mined mass (metric ton) for 39 chemical elements showing a power-law relationship.  A similar power law relates price to crustal abundance. [Jonathan Price, personal communication]}
    \label{fig:zprice}
\end{figure}

Jonathan Price, the Nevada State Geologist and cochair of the ECE study group, compiled the price-abundance relationship for most of the natural chemical elements \cite{ECE}.  Remarkably, a somewhat noisy power-law relationship between price and crustal abundance unmistakably appears.    After converting crustal abundance to mined mass, the log-log plot in Figure \ref{fig:zprice} shows that despite the noise a power-law pattern with a kink covers several decades; this remarkable finding prompted us to seek its underlying cause as a path to inform materials criticality assessment.  Price and abundance (or mined mass) are just two factors in supply risk, and tying them together with a mechanism eliminates one variable; there are many such variables and others may yield to an entropic analysis.

Entropic approaches have been explored before by economists.  A price entropy was investigated by Yakavenko and Rosser \cite{yakavenko} who applied it to supply-and-demand economics.  The underlying assumption is that markets explore many possible configurations through the choices of many agents from mining engineers to investors.  Several variables chosen under market forces would be candidates for entropic arguments.  Here we apply entropy to geology by exploring how price adjusts to crustal abundance.

Figure \ref{fig:zprice} displays prices $\{p_1, p_2,...p_i,...p_N\}$ against mined mass which in turn maps to crustal abundances $\{n_1, n_2,...n_i,...n_N\}$ where $N\approx100$ is the number of elements (or classes of elements).  Either set may be chosen to construct an entropy function; we choose $n_i$ here.  The goal is to vary the distribution of $n_i$ with constraints to find the optimal variation of price.

Our fundamental assumption is that the distribution of values $\{n_i\}$ is strongly peaked around the most probable one.  A clarifying way to look at the distribution function $W(n_i)$ is to consider an ensemble of regional geologies and define the entropy in the number of accessible (abundance) states $\Omega$ as

\begin{align}
{S}_m = -k \ log\ \Omega  .
\label{entropybasis}
\end{align}

Although $k$ is normally Boltzmann's constant, given that we are dealing with geologic as opposed to thermodynamic quantities, we delay assigning a value to it since it gets combined with integration constants.  

Equation \ref{entropybasis} can be quickly rationalized and put into useful form following Landau and Lifshitz Statistical Physics \cite{Landau}.

\begin{figure}
    \centering
    \includegraphics[width=0.7\linewidth]{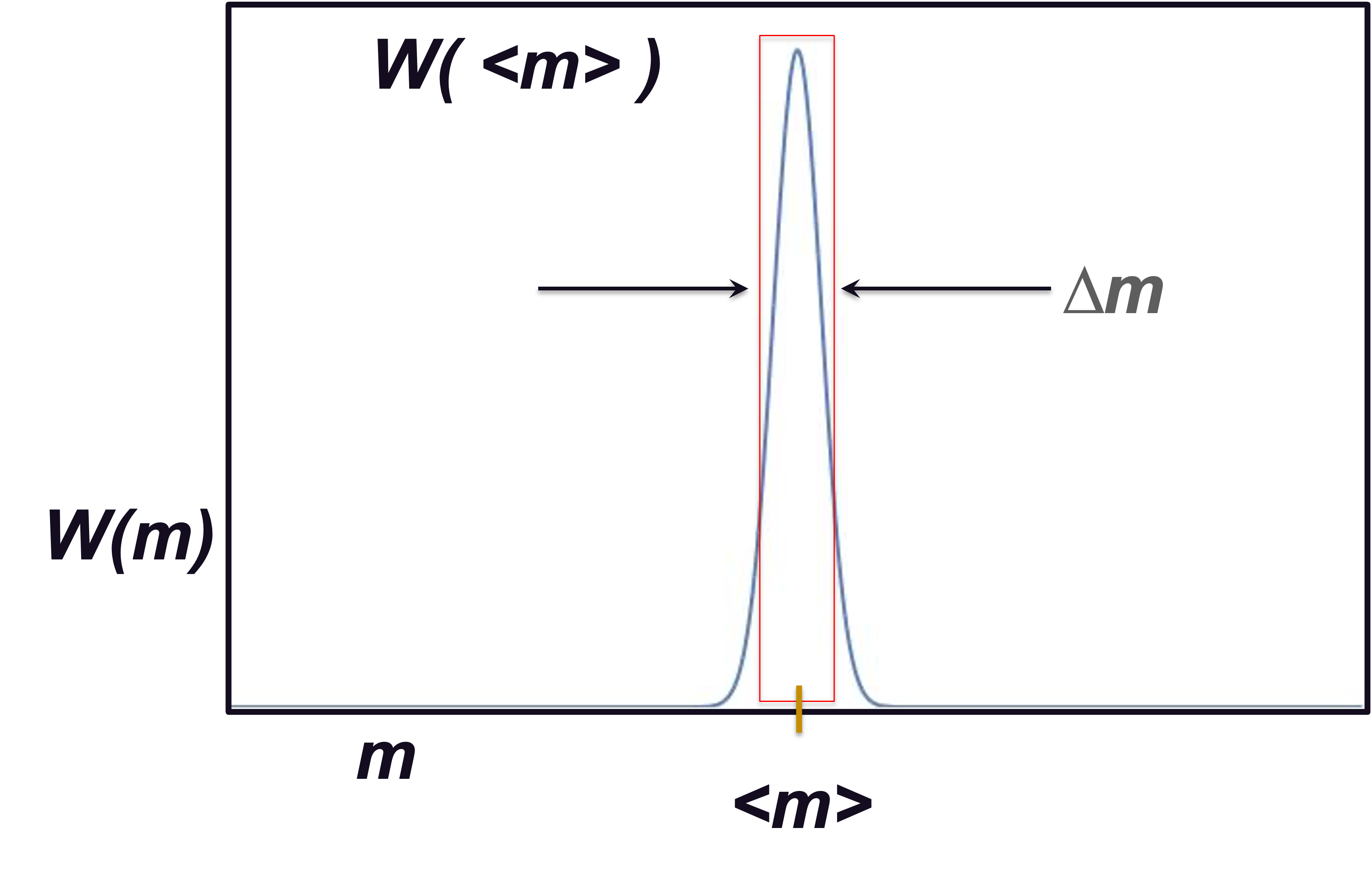}
    \caption{Expected distribution of element masses $m_i$ in an ensemble of regional geologies}
    \label{fig:zdistribution}
\end{figure}

In the phase space of $\{n_i\}$ or equivalently the (extracted) masses $\{m_i\}$, the density of states $\rho(m)$ can be expected to be sharply concentrated at a mean mass $\langle{m}\rangle$ resulting in a sharp peak in $W(m)$ of a narrow width $\Delta m$ as shown in Figure \ref{fig:zdistribution}.

For simplicity of normalization, the box enclosing the peak of $W(m)$ has width $\Delta m$.  With $M$ defined as the total mass of all elements, the normalization of $W(m)$ and the average mass $\langle m\rangle$ are

\begin{align}
    W(\langle m\rangle) \Delta m &\equiv 1 ,\\
    \langle m\rangle &= \sum \limits_{i=1}^{N} m_i n_i ,\\
    M &= N \langle m\rangle  .
\end{align}

The next step in this development is to relate the number of accessible states $\Omega$ to the density of states $\rho(m)$ via its definition,

\begin{align}
    \rho(m) dm \equiv \frac{d\Omega}{dm} dm
\end{align}

At the mean mass $\langle m\rangle$, the volume of phase space $\Omega$ can be expressed as an arbitrary baseline $\Omega_0$ less a differential volume $\Delta \Omega$,

\begin{align}
    \Delta \Omega = \Omega(\langle m \rangle ) - \Omega_0 
\end{align}

Combining results,

\begin{align}
{S}_m &= k \log \Delta \Omega\\
&= k \log \frac{1}{n( \langle m \rangle )}\\
&=-k \log n( \langle m \rangle )\\
&=-k \langle \log n(m) \rangle \\
&=-k \sum \limits_{i=1}^{N} n_i \log n_i
\label{finalentropy}
\end{align}

Equation \ref{finalentropy} is the starting point for finding an extremum using a variational method.  The entropy $S_m$ is not very useful by itself until we add constraints on total number of elements and on real-world economics.  Before adding economics, it is useful to pause to consider the meaning of $S_m$ for which the distribution of elemental abundances is still unconstrained.

First, consider the case in which the geology consists of only one element $n_0=1$.  Equation \ref{finalentropy} yields $S_m=-k n_0 \log{n_0}=0$.  This geology could be considered the most ordered possible.  Perhaps the only realization of this case is a neutron star!

What would be the least ordered geology?  We apply Lagrange multipliers \cite{lagrange} with a constraint on the number of elements

\begin{align}
    \sum \limits_{i=1}^{N} n_i = 1 ,
    \label{totalnumber}
\end{align}

to maximize the entropy yielding

\begin{align}
    n_i &= 1/N\\
    S_m &= N k \log{N} .
\end{align}

The least ordered geology has equal fractions of elements in the distribution.

While other simple geologies can be explored, a basic economic constraint is more important to study, namely, the constraint of finite budget.

\begin{align}
     \sum \limits_{i=1}^{N} m_i p_i = T ,
     \label{totalmoney}
\end{align}

where the constant T is the total money available for mining, extraction, refining, and processing.  

To find an extremum of $S_m$ subject to constraints Equations \ref{totalnumber} and \ref{totalmoney} we use the method of Lagrange multipliers \cite{lagrange}.  The extreme value occurs when small variations vanish of the quantity

\begin{align}
     U_m = S_m + \alpha N + \beta T ,
     \label{utility}
\end{align}

where $U_m$ is the utility function and $\alpha$ and $\beta$ are Lagrange multipliers to be determined.  The subscript $m$ reminds us that we have chosen to seek the extreme value by varying the distribution $\{m_i\}$ or equivalently $\{n_i\}$.

\begin{align}
     0 &= \delta (U_m) \\
       &= \delta (S_m + \alpha N + \beta T) \\
       &= \sum \limits_{i=1}^{N} [( -k (1 + \log n_i ) \delta n_i + \alpha m_0 \delta(n_i p_i) + \beta \delta n_i ]
     \label{steps}
\end{align}

Using the fact that the variations $\delta n_i$ are independent and small, yields a simple differential equation and solution for price $p(n)$

\begin{align}
    \alpha m_0 (p(n) + \frac{dp(n)}{dn}) = -\beta + k ( 1+ \log n )\\
    p(n) = \frac{c}{n} + \eta (\log{n} - \lambda)
    \label{perumalcase1}
\end{align}

where $c$ and $\eta$ are constants determined by fitting to the data.  Equation \ref{perumalcase1} immediately shows $p$ scales as $n^{-1}$ as noted in the data for small $n$.  The fit to 39 elements is shown in Figure \ref{fig:zprice}.

To be able to relate price to crustal abundance or mining production is helpful in cases when criticality assessment turns to qualitative analysis for lack of quantitative data.  It is also important to find underlying mechanisms in markets for greater understanding of criticality behavior.  While the proposed entropic quantities are random variables in the sense that they derive from an entropic probability distribution, they are not dynamic predictors over time; with deeper study, for example examining price data time series, it may be possible to explain transitions, such as price shocks or evolutions in criticality.  For the present, the purpose is to build on prior work \cite{yakavenko} to see if underlying market mechanisms can be identified.

\subsection{Acknowledgments and Funding}\label{acknowl}

This paper would not have been possible without significant intellectual contributions by R. Perumal Ramasamy of Anna University, my coauthor of reference \ref{perumalcase1} summarized in part in this paper.  This review arose from the Materials Research Society Spring 2025 Meeting, I wish to thank my colleague and frequent co-author Min-ha Lee of Stanford University and KITECH who organized the symposium for which I was designated a Distinguished Speaker.  This work was supported in part by the US Department of Energy through the Los Alamos National Laboratory. Los Alamos is operated by Triad National Security, LLC, for the National Nuclear Security Administration of U.S. Department of Energy (Contract No. 89233218CNA000001).

\bibliography{sn-bibliography}

\section{Contributions by Authors}\label{contrib}

Only the sole author AJH contributed to this work.  In the Entropic Approach section, collaborator R. Perumal Ramasamy's contribution is cited in reference \ref{perumalcase1} appropriately; the extension of this section to develop geologic entropy--not reported in \ref{perumalcase1}--is by AJH.

\section{Conflicts of Interest}\label{COI}
Not applicable.

\section{Data Availability}\label{websiteslist}

Table \ref{tab:websites} lists websites used in this study and dates when they were accessed to check availability.  This list is not intended to be comprehensive as there are many more online sources for mineral and elemental information.

\backmatter

\begin{appendices}

\begin{table}[ht]
    \centering

\begin{tabularx}{1.0\textwidth}  {|p{1.1cm}|p{3.6cm}|p{5.4cm}|p{1.2cm}|}

 \hline
 \multicolumn{4}{|c|}{\textbf{Web Resources for Critical Elements}} \\
 \hline
 Elements& Institution & Website & Accessed \\
 \hline
 all   & U.S. Geological Survey, Mineral Commodity Summaries    & $https://www.usgs.gov/centers/-national-minerals-information-center$ & 2Sept2025\\
 all &   CRC Handbook of Chemistry and Physics, 97th ed  & $https://mathguy.us/BySubject/Chem-istry/CRC-Handbook-of-Chemistry-and-Physics-97.pdf$   &2Sept2025\\
 He &Sigma-Aldrich & $http://www.webelements.com/helium/$ & 2Sept2025\\
 C, U    &US Energy Information Administration & $https://www.eia.gov/nuclear/$ & 2Sept2025\\
 Nd    & Statista: Asian Metal, Stormcrow & $https://www.statista.com/outlook/-io/mining/base-metals/asia$ & 2Sept2025\\
 
 REE&   Argus Media (formerly metal-pages.com)  & $https://www.argusmedia.com/en/com-modities/metals$ & 2Sept2025\\
 Sm & Inframat Advanced Materials  & $http://www.advancedmaterials.us/-62R-0801.htm$   & 2Sept2025\\
 U & World Nuclear Association  & $http://www.world nuclear.org/info/-inf23.html$ & 2Sept2025\\
 U & Costmine Intelligence  & $http://www.infomine.com/-investment/historicalcharts/-showcharts.asp?c=uranium$ & 2Sept2025\\
 \hline
\end{tabularx}

    \caption{Web resources for elements used in  this study. REE stands for rare earth elements.}
    \label{tab:websites}
    
\end{table}

\end{appendices}

\end{document}